# A Spatial Knowledge Acquisition Comparison Between Digital Visual Thematic Maps, Non-Visual Interactive Text Thematic Maps, and Tables

BRANDON BIGGS (CONTACT AUTHOR): BRANDON.BIGGS@SKI.ORG – GEORGIA INSTITUTE OF TECHNOLOGY AND THE SMITH-KETTLEWELL EYE RESEARCH INSTITUTE

CHRISTOPHER TOTH: CHRISTOPHER.TOTH@XRNAVIGATION.IO – XR NAVIGATION

JAMES M. COUGHLAN: COUGHLAN@SKI.ORG – THE SMITH-KETTLEWELL EYE RESEARCH INSTITUTE

BRUCE N. WALKER: BRUCE.WALKER@PSYCH.GATECH.EDU – GEORGIA INSTITUTE OF TECHNOLOGY

Abstract

Digital maps are used to communicate generalized spatial information and relationships, yet are commonly made “accessible” using tables that lack geographic information. This study examines whether these tables and interactive text maps (ITMs) may be comparable to visual maps. Twenty sighted and 20 blind and low-vision individuals (BLVIs) performed tasks designed to compare visual maps, ITMs, and tables. Participants answered numeric, geographic, and combined numeric geographic questions using each representation, and performance, preference, and NASA-TLX were measured.

Across both participant groups, map representations (visual and ITMs) significantly outperformed tables on geographic-based questions, while performance differences were minimal for numeric questions. For sighted participants, performance on geographic questions did not significantly differ between visual maps and ITMs, indicating that a larger powered study may find an “equivalent purpose” across these two conditions. Participants preferred map-based representations over tables. Perceived workload was highest for the ITM, intermediate for the visual map, and lowest for the table.

Consistent with the Map Equivalent Purpose Framework, these findings indicate that Web Content Accessibility Guidelines-compliant ITMs can provide access to spatial information, unlike tables. These findings challenge prevailing accessibility practice that recommends tables lacking geographic information as map alternatives, and motivate reconsideration of accessibility legislation exempting digital thematic maps.

## 1 COMPUTING CLASSIFICATION

- **Human-centered computing** / *Empirical studies in accessibility*; *Accessibility technologies*; *Accessibility design and evaluation methods*; *Empirical studies in HCI*; *Geographic visualization*.
- **Applied computing** / *Cartography*.

## 2 INTRODUCTION

Cartography and geographic maps have existed in paper form for thousands of years, but since the mid-1960s, maps have begun to transition to digital formats with the advent of graphic displays [Longley et al., 2011a]. Digital maps are now ubiquitous and affect every aspect of daily life, from transporting goods around the world to displaying election results or tracking hurricanes [Longley et al., 2011b]. Despite the prevalence of maps, none of the 285 million blind and low-vision individuals (BLVIs) have easy access to digital geographic maps [World Health Organization, 2010], and BLVIs typically view fewer than one tactile map per year [Biggs, Pitcher-Cooper, et al., 2022; Giudice, 2018; Juan-Armero and Luján-Mora, 2019]. In contrast, sighted individuals view more than 300 maps per year [Savino et al., 2021]. This represents a massive gap in knowledge access that must be closed.

There are two types of maps: referential maps, in which the geographic features are paramount, and thematic maps, where numeric or categorical variables are overlaid on geographic features [Longley et al., 2011c]. Past research on non-visual maps has primarily focused on referential maps used for navigation [Biggs et al., 2019; Brock et al., 2015; Ducasse et al., 2018; Giudice et al., 2020; Ottink et al., 2022]. However, this paper focuses on thematic maps, in particular choropleth maps, for three reasons: (1) Many cartographic dependent professions (e.g., oceanography, epidemiology, meteorology, climate science, geology, city planning, etc.) primarily use *thematic choropleth* maps [Department of Geography and Geosciences, Salisbury University; Longley et al., 2011b]. With BLVIs having higher unemployment and poverty rates than the population as a whole [O'Donnell, 2014], there is a clear need to make more professions accessible to BLVI workers, and this includes professions that rely on thematic maps; (2) Unfortunately, major digital accessibility legislation actually makes this employment goal harder to achieve by having explicit exceptions to accessibility requirements for thematic maps, due to the perceived intractability of making maps accessible [European Commission, 2017; The public sector bodies (websites and mobile applications) (no. 2) accessibility regulations 2018, 2018]; (3) This paper focuses on accessibility options for thematic choropleth maps. choropleth maps present the clearest case for evaluation because each geographic feature has exactly one thematic value showing on the map at once, avoiding the complexity of overlapping symbols (e.g., election result maps, disease maps, etc.) [Types of thematic maps, 2017].

Choropleth maps typically use different colors or shades to visually represent different values of a variable (e.g., the number of cases of an illness like COVID-19), and each feature (e.g., each state in a USA-based map) only has one value at a time (see Figure 1 below).

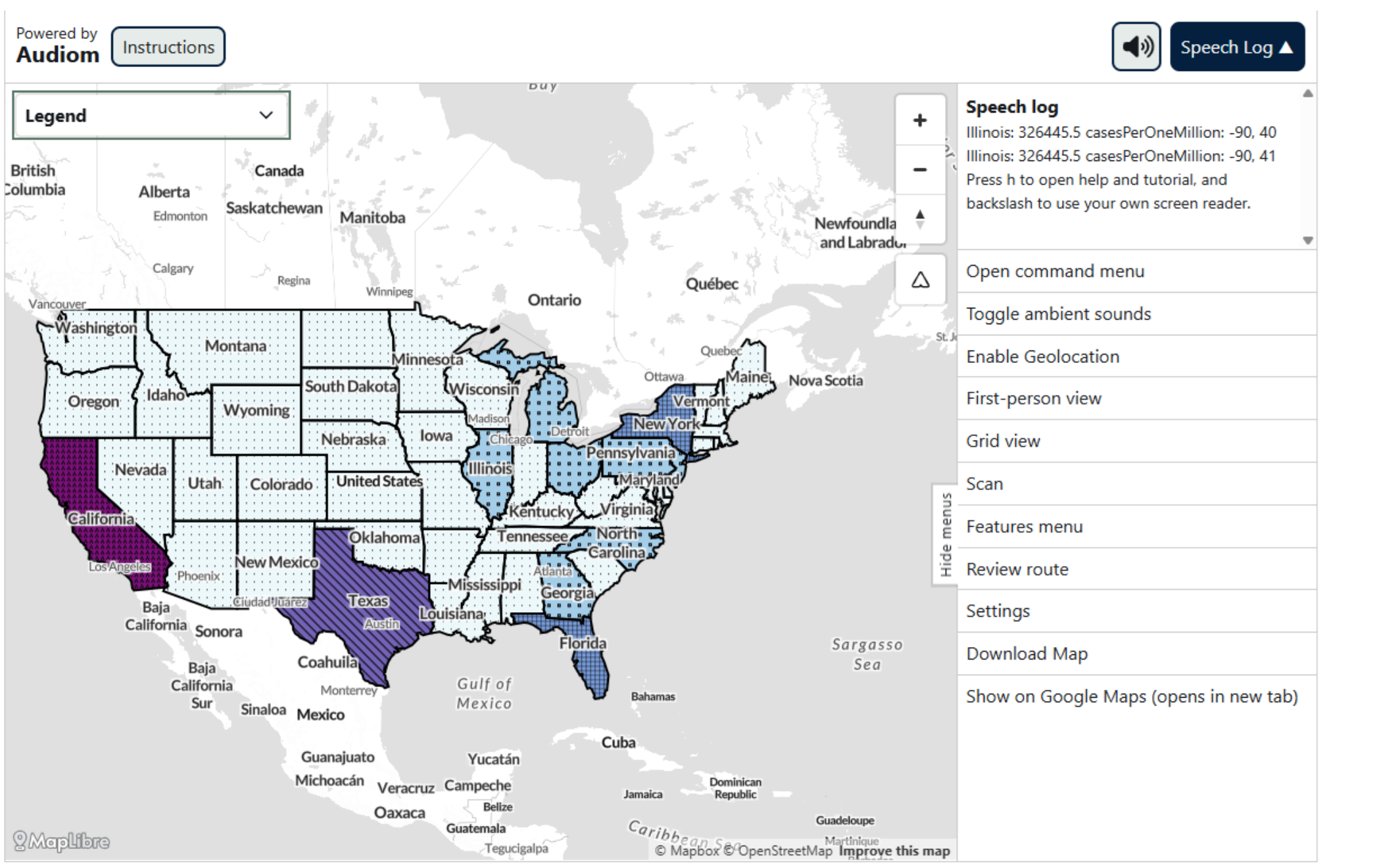

Figure 1: A visual image of a choropleth map showing different colors over different geographic features (states) over the U.S. The colors represent different numbers of cases of COVID-19 in that state.

Note: An interactive Audiom example of this U.S. COVID statistics choropleth map can be found at this link (Opens a New Tab).

Approximately 16% (30 million) of publicly facing websites have digital maps [Bocoup, 2022]. Furthermore, the digital map market was valued at around $28.5 billion in 2024 [Digital map market, 2019], and over 20 professions (e.g., meteorology, oceanography, epidemiology) depend on digital choropleth maps [About and mission]. Without access to digital geographic maps, BLVIs are excluded from civic engagement, numerous professions, educational courses, news articles, and public health information. Although navigation map evaluations are more common, thematic maps are required for professional cartographic work [Longley et al., 2011b]. This paper extends the thematic map evaluation on the interactive text map (ITM) interface, Audiom, which was used in this study [Biggs, Toth, et al., 2022].

Blind individuals generally use screen reader software to access digital content, including maps. Unfortunately, and frustratingly, when it comes across a traditional visual map, a screen reader will typically say the word "graphic", which is clearly of no use to the blind user. Also, low-vision users are generally unable to customize visual maps for their needs, such as changing colors or contrast [Biggs, Toth, et al., 2022; Biggs et al., 2025]. So maps pose a problem for both blind and low-vision users. In a recent evaluation of the top 14 digital map tools (e.g., Google Maps, Apple Maps, ESRI), only Audiom met even the most basic level of the Web Content Accessibility Guidelines (WCAG) [Biggs et al., 2025; World Wide Web Consortium]. All 13 failing tools read out "graphic blank" to screen readers attempting to access geometries and were impossible to fully use with the keyboard.

The Map Equivalent Purpose (MEP) Framework quantifies equivalency between visual maps and text maps using the MEP variables including shape, size, orientation, distance, direction, topological relationships, and location information [Biggs et al., 2026]. An evaluation of eight text maps indicates that ITMs and well-crafted audio descriptions can convey all the information present on visual maps. Unfortunately, while these methods for effectively representing spatial information non-visually have been developed in research settings [Ducasse et al., 2018; Ottink et al., 2022], major digital map companies have not yet adopted these techniques [Biggs et al., 2026; Chan and Linder, 2021]. Instead, the usual method for attempting to make digital choropleth maps "accessible" to BLVIs remains providing data tables [Biggs et al., 2026; Sloan, 2020], even though tables may not convey the spatial information present in the (visual) map. Therefore, it is necessary to systematically compare digital visual maps, ITMs, and tables using the MEP framework. To our knowledge, this is the first such assessment [Biggs et al., 2026; World Wide Web Consortium].

### 2.1 Related Work

Tactile maps are the traditional non-visual map representation modality, but their lack of availability, high literacy requirement in both braille and tactile graphics, static nature, expense, lack of standardization, and required simplification mean they are inadequate as a mainstream digital map interface [Biggs, Pitcher-Cooper, et al., 2022; Butler et al., 2017; Ducasse et al., 2018; Longley et al., 2011c; Rowell and Ongar, 2003; Rowell and Ungar, 2003, 2005; Weimer, 2017]. Work has been done to make tactile maps interactive which reduces braille literacy requirements, allows dynamic numeric and categorical variables, and allows for greater feature density, but all the other above limitations remain [Brock and Jouffrais, 2015; Ducasse et al., 2018; Landau and Gourgey, 2001; Manzoni et al., 2025; Papadopoulos et al., 2018]. New full-page tactile displays have recently entered the commercial market that may possibly allow larger dynamic tactile maps, dynamic numeric and categorical variables, and greater feature density, but they still suffer from all other previously mentioned limitations (with the addition of a massive price tag) [American Printing House for the Blind]. Full-page braille displays also lack keyboard accessibility, and are not text-based, so are unable to act as a WCAG compliant solution [Biggs et al., 2025]. The limited research on full-page braille displays is promising, although the mentioned downsides will probably keep these displays from becoming a mainstream ubiquitous solution in the near future [Rao and O'Modhrain, 2020]. Because tactile maps are not considered a WCAG compliant solution, they were not evaluated as a condition in this study. There has been no research on full-page tactile displays and thematic maps. In contrast, the following non-visual digital map displays are usable on common phones, tablets, or both.

Using a taxonomy of input modalities, digital non-visual thematic map research can be categorized into both interactive text maps and static audio descriptions. Interactive text maps can be split into touchscreen and keyboard. It is critical to note that if an interactive map is to be considered WCAG compliant, all information needs to be presented through text messages sent to the screen reader with audio being supplemental to the text, creating an ITM [Biggs et al., 2025; World Wide Web Consortium].

Touchscreen non-visual maps use speech, vibration, and audio as users move their finger over geographic features [Carroll et al., 2013; Delogu et al., 2010; Ducasse et al., 2018; Ottink et al., 2022]. These interfaces have been shown to convey direction and exact location well but struggle with precise distance, orientation, shape, and size [Delogu et al., 2010; Ottink et al., 2022]. Critically, WCAG requires keyboard accessibility, which eliminates touchscreen-only maps from being a WCAG compliant non-visual digital map solution [Biggs et al., 2025; World Wide Web Consortium].

Keyboard ITMs require users to move an avatar around a map, typically jumping between features using arrow keys, using speech to identify feature properties, playing sounds representing values on that feature, and have other key shortcuts to query information about the map [Delogu et al., 2010; Ducasse et al., 2018; Zhao et al., 2008]. Although most keyboard map applications have been built for Microsoft Windows due to its overwhelming popularity among BLVIs, their conventions can be used on any keyboard enabled device, including smartphones with a Bluetooth keyboard, which make them ideal for WCAG compliance and for potential broad adoption and integration with mainstream digital map applications [Biggs et al., 2025; Screen reader user survey 10 results, 2024; World Wide Web Consortium]. BLVI screen reader users already operate the computer exclusively with a keyboard (either braille or QWERTY), so a keyboard interface is most intuitive and pleasant to use [Kreimeier et al., 2020; NV Access, 2017]. Keyboard maps have been shown to effectively convey all MEP variables across a large number of (over 50) points, polygons, and lines [Biggs et al., 2019; Biggs, Toth, et al., 2022; Delogu et al., 2010; Ducasse et al., 2018; Ottink et al., 2022]. Delogu et al. and Zhao et al. developed an interface that displays thematic choropleth maps of U.S. population data by combining a table view with a grid that shows each state, allowing users to navigate with arrow keys while hearing a pitch and spoken data for each state [Delogu et al., 2010; Zhao et al., 2008]. In an evaluation with seven BLVIs, Zhao et al. found that users answered geographic questions with 95% accuracy using this interface compared to only 20% accuracy with an Excel table, though they reported difficulties such as the absence of geographic shapes, challenges with variable correlation, and an average tutorial time of 109 minutes [Zhao et al., 2008]. Later, Delogu et al. extended the evaluation to 20 BLVI and 10 blindfolded sighted participants who, in four trials, rated the similarity between digital and tactile maps with a mean of 7.3 versus 3.75 ($p < .01$) and showed no performance difference between keyboard and touchscreen input [Delogu et al., 2010]. There was an interface from Guerreiro et al. that did use a touchscreen interface to control an avatar as a mixture between a touchscreen and keyboard interface, and distance, direction, and general layout of the map was communicated [Guerreiro et al., 2017]. Other digital keyboard map research focuses on referential maps (e.g., Loeliger et al. [Loeliger and Stockman, 2014], Heuten et al. [Heuten et al., 2006], Heuten et al. [Heuten et al., 2007], Picinali et al. [Picinali et al., 2014], Connors et al. [Connors et al., 2014]) but there is enough research in thematic keyboard maps to not discuss these interfaces in depth here. Most referential interfaces focus on using looping spatial audio around the user's head to convey the location of features, but this convention has rarely been used in thematic map research. The only example is a brief mention of using looping sounds of decreasing wildlife activity to represent the increasing level of totality on an eclipse isopleth map showing lines of totality over the United States [Walmer et al., 2024].

Audio descriptions of maps (or detailed text descriptions) typically consist of a data table with geographic information, a tree structure of feature values, or a long text description [Hennig et al., 2017; Smelcer and Carmel, 1997; Zong et al., 2022]. Text can be output visually, auditorily through a screen reader, and tactilely through a braille display [Biggs, 2024; Web Content Accessibility Guidelines (WCAG) overview, 2018]. Evaluations have found high cognitive load and limited spatial knowledge transfer for audio descriptions [Hennig et al., 2017; Ottink et al., 2022; Zong et al., 2022], though BLVIs have demonstrated understanding of distance, direction, and exact location from text descriptions [Schmidt et al., 2013]. The UniDescription project provides guidelines for map descriptions used by the National Park Service, but no quantitative spatial evaluation has been performed [Conway et al., 2020; UniDescription]. The MEP Framework evaluation found that audio descriptions can communicate all spatial information if carefully created, but no user evaluation has validated this [Biggs et al., 2026]. The need for manual production, static nature, high maintenance cost, and inherent requirement for a separate page lead this project to focus on interactive alt-text keyboard maps.

One prior evaluation compared visual maps with tables, although those tables included a column in the table listing the bordering geographic features [Smelcer and Carmel, 1997 ]. The study by Smelcer et al. [Smelcer and Carmel, 1997] evaluated only distance and found that visual maps were significantly better than complex tables containing spatial descriptions of adjacent features. This suggests that even if simple tables provided to BLVIs as an alternative to a thematic map did contain geographic information (which they do not), they would still be less effective than any kind of geographic map.

Repeated studies over the years have demonstrated BLVIs are able to build cognitive maps undistinguishable from sighted participants [Golledge et al., 2000; Goodridge et al., 2021; Ottink et al., 2022]. This mental map by BLVIs is often ego-centric and created sequentially [Giudice, 2018; Golledge et al., 2000; Iachini et al., 2014]. Ego-centric Mental maps are most effectively built up by people moving themselves around the environment, creating a map from sequential information following Embodied cognition theory [Burgess et al., 2002; Newcombe, 2024]. Routes between features can be placed into a larger gestalt understanding of space, and being able to connect routes, through viewing a map representation, provides a stronger cognitive map than only navigating routes [Jacobson, 1998; Siegel and White, 1975; Zhang et al., 2014]. ITMs utilize these cognitive map theories, allowing users to control an ego-centric avatar or finger, and sequentially construct a mental map as they move their avatar or finger around the space [Biggs, Toth, et al., 2022].

## 3 METHODS

### 3.1 Platform Design

The Github repository with code for all map conditions can be found in the metadata.

#### *3.1.1Interactive Text Map*

The ITM platform used in this evaluation is Audiom, a web-based keyboard map viewer previously evaluated in [Biggs, Toth, et al., 2022] and co-designed in [Biggs et al., 2024]. Participants used only the audio output; braille display and visual text modes were not used. Audiom has two connected navigation modes: a geographic view and a tabular view. In the geographic view, users move an avatar a user-specified distance using arrow keys (e.g., up arrow moves north). As the avatar crosses a feature, a pitched tone conveys the statistical value (from a two-octave range), followed by a text message with the feature name and exact value (e.g., "California, 1346 cases per 1 million") [Krygier, 1994; Walker and Mauney, 2010; Walker and Nees, 2011; Zhao et al., 2008]. Since all information is accessible through text, Audiom is classified as an ITM [Lee et al., 2022]. The geographic view employs an egocentric perspective, which is more understandable to BLVIs than the allocentric perspective of visual maps [Biggs et al., 2018, 2019; Giudice, 2018; Giudice et al., 2020; Iachini et al., 2014]. In this evaluation, the avatar orientation was locked facing north. In usability testing with 55 BLVIs, the keyboard interface was highly preferred and quick to learn [Biggs et al., 2019, 2024; Biggs, Toth, et al., 2022], and the MEP Framework evaluation found it was one of three text maps that showed equal access to a visual baseline [Biggs et al., 2026].

The Audiom choropleth interface was evaluated for WCAG compliance by LevelAccess [LevelAccess.com: ADA Compliance, 508 Compliance, WCAG, VPAT], a 3rd-party digital evaluation company, and their publicly available Accessibility Conformance Report (ACR) showed that Audiom meets WCAG AAA compliance [XR Navigation], the highest level of accessibility compliance [Web Content Accessibility Guidelines (WCAG) overview, 2018]. Audiom seems to be the first digital map tool to have ever met WCAG A compliance [Biggs et al., 2025]. Audiom also seems to be the only academically evaluated interface with an ACR, despite the legal importance of having an ACR. This means that results from this evaluation will be directly applicable to digital accessibility and legislative action.

The map data powering Audiom is standard GeoJSON, the data powering many maps on the web [Working Group Geographic JSON (geojson), 2016]. GeoJSON is an inherently amodal data format, as there is no requirement for visual only properties. Using GeoJSON allows Audiom to subscribe to a representation-agnostic data format as described in Zong et al. [Zong et al., 2024]. The only modification that Audiom requires to the GeoJSON specification, is a "name" property to be added to each feature, describing what it is (e.g., "California", or "Wall") [Biggs, 2020].

*3.1.2 Visual Map*

Although Audiom does have a visual map, it was turned off during this evaluation, as at the time it did not have strong support for custom visual geometries (Audiom now does support these custom geometries). The custom visual map was built using Mapbox [Mapbox, 2020], and React [React: A JavaScript library for building user interfaces, 2018]. The visual map did not meet all the WCAG criteria, as some of the contrast and hover requirements would have required significant Mapbox modification, but this was not seen as a significant factor, since no Mapbox maps meet these WCAG criteria [Biggs et al., 2025]. Since the sighted participants were required to have 20/20 vision, using a "standard" visual map was deemed sufficient. Participants could pan using the mouse, open a data table with one column for the state name, and another column for the statistical values, and switch statistics using a combo box. The statistical values were represented using different color shades. Each statistic had a different color when the user switched to it, and a lighter color meant higher cases, and a darker color meant lower cases. There was a legend that showed the numeric segmentation colors, like a traditional choropleth map.

*3.1.3 Table Map*

The table map was built using the table component from Material UI [Material UI SAS, 2026], with full keyboard support. Each row of the table showed the statistical values for each feature on the map. The left most column contained the feature name, and the columns to the right contained the statistical values associated with those features. Activating (with the mouse or keyboard) the header of each column would sort the rows by that column through three options: Ascending, descending, or not sorted. This table was intentionally meant to be similar to the tables used as a text alternative at locations like the CDC or Esri [McCall, 2024; United states COVID-19 cases, deaths, and laboratory testing (NAATs) by state, territory, and jurisdiction]. A column for geographic information was intentionally excluded because most tables used as a map alternative lack a geographic column.

### 3.2 Study Design

*3.2.1 Research Questions*

The overarching research question for this study (RQ1) was: How does spatial knowledge acquisition compare between a thematic interactive text map (ITM), a thematic visual map, and a table? This study investigated performance, preference, and subjective responses within sighted and BLVI participants when answering spatial questions using these three modalities. The dependent variables were performance, preference, and subjective ratings (NASA-TLX), while the independent variable was presentation modality (visual map, ITM, and table). Performance within groups was the primary focus, with no between-group comparisons conducted.

RQ1 was decomposed into the following sub-research questions, each corresponding to an analysis reported in the Results section:

- RQ1.1 (Performance by Condition): How does performance compare within sighted and blind participants when answering spatial questions on the visual map, ITM, and table?
- RQ1.2 (Performance by Question Type): How does performance within blind and sighted participants compare across the different question types (Spatial Knowledge Numeric, Spatial Knowledge Geographic, and Spatial Knowledge Numeric Geographic) across the different map conditions (visual map, ITM, and table)?
- RQ1.3 (Visual vs. ITM on Geographic Questions, Sighted): How does performance on Spatial Knowledge Geographic questions compare between the visual map and ITM conditions within sighted participants?

- RQ1.4 (Subjective Confidence): How do self-reported confidence levels vary within blind and sighted participants when answering spatial questions using the visual map, ITM, and table?
- RQ1.5 (NASA-TLX Cognitive Workload): How does the NASA Task Load Index compare within blind and sighted participants across the visual map, ITM, and table conditions?
- RQ1.6 (Preference): How do condition preferences vary within blind and sighted participants across the available presentation modalities?
- RQ1.7 (Within-Subject Correlations): How consistent is within-subject performance across conditions for blind and sighted participants, and does this consistency differ by question type?

#### *3.2.2Spatial Knowledge Evaluated*

This evaluation assessed several MEP variables. Direction was evaluated through explicit questions about feature locations (e.g., identifying features west of another feature or along the east coast). Topological relationships were assessed through questions about bordering features and coastline objects. Although shape, size, orientation, and distance were not explicitly measured, they were implicitly evaluated as participants needed to explore irregular features of varying sizes and orientations to identify all bordering features. Exact location was assessed through questions requiring identification of features in specific regions (e.g., northwest corner, east coast, land-locked areas). Spatial relationships between thematic variables were explicitly evaluated through numeric questions (e.g., identifying features based on combined spatial and thematic criteria). Only polygon features were used in these maps. Future work should explicitly measure shape, size, orientation, and distance variables.

#### *3.2.3Data*

Four fictional maps were used in the study: one training map and three experimental maps. Each map consisted of twenty irregular fictional features (i.e., reminiscent of a US state like Nevada), arranged within a rectangular region (i.e., reminiscent of the 48 contiguous US states) with variations in shifting, rotation, names, and numeric values (see figure 2 for an example of one of the four maps). The geometries were intentionally made as similar to a real-world epidemiological choropleth map as possible [United states COVID-19 cases, deaths, and laboratory testing (NAATs) by state, territory, and jurisdiction]. The 80 fictional town names were generated using a town name generator and were based on real towns to ensure ease of pronunciation, as verified by pilot participants. Random statistical values for each feature were generated in Excel using the RANDBETWEEN formula. Participants viewed fictional COVID-like statistics for Total Cases, total deaths, and total tests for each feature. Other columns with random values were included but not used in the study.

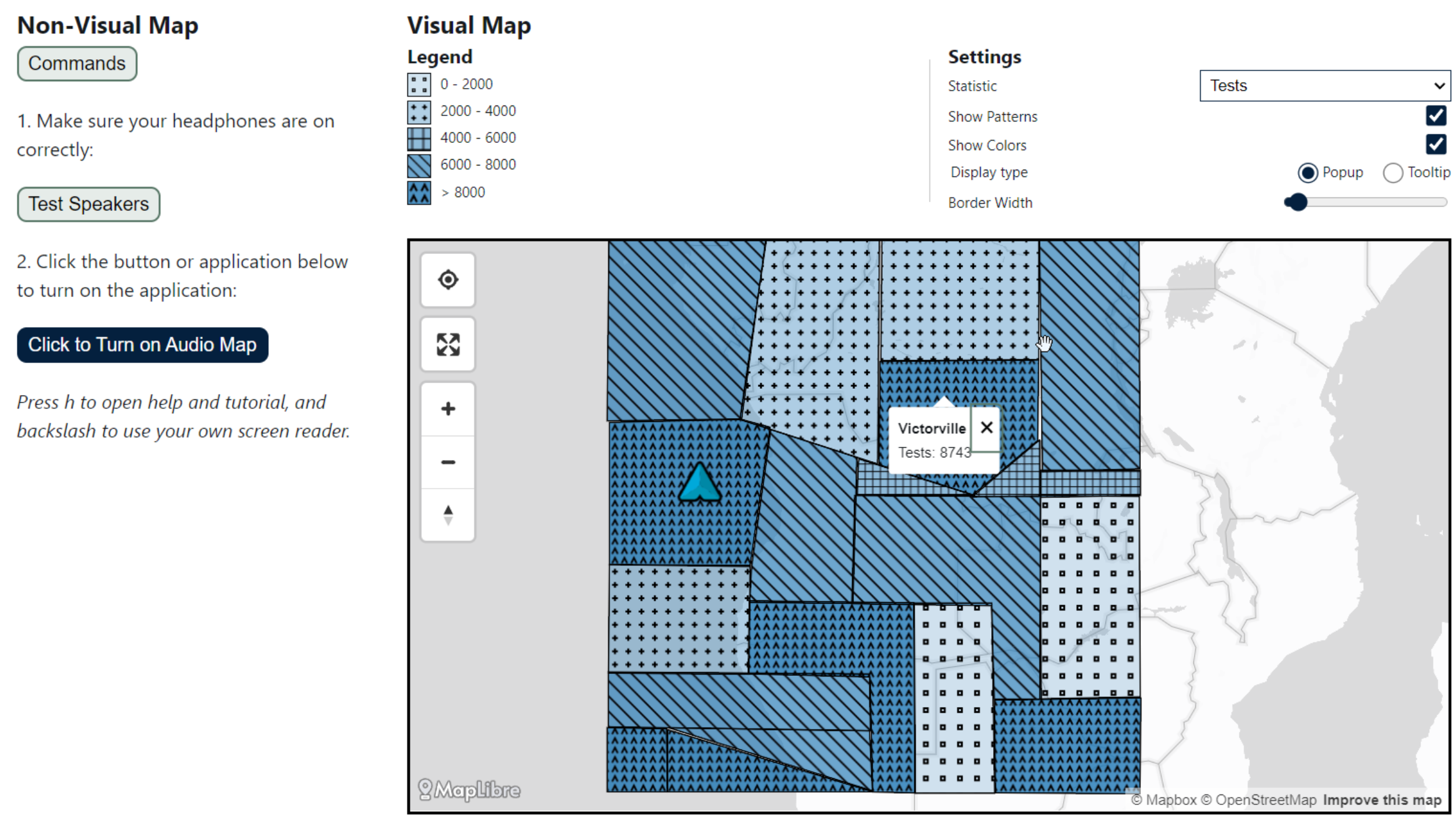

Figure 2: A picture of a choropleth map of 20 irregular polygons showing statistical values
Note: An interactive Audiom map of this test condition can be found at this link (Opens a New Tab).

*3.2.4Participants*

Forty remote participants were recruited for the study, including 20 blind and 20 sighted individuals. BLVIs ranged in age from 21 to 68 years old, and included 12 females and 8 males. Sighted participants ranged from 18 to 24 years old, and included 12 females and 8 males. BLVIs were recruited through email lists and participant databases, while sighted student participants were recruited through the student participant recruitment system at Georgia Tech. BLVIs included 2 people who identified as Hispanic or Latino, and 6 Asian, 10 White, 1 American Indian/Alaskan Native, 1 Mexican, and 1 Black or African American. Sighted participants included 2 participants who identified as Hispanic or Latino, and 10 Asian, 7 White, 1 Black or African American, 1 American Indian/Alaskan Native, and 1 no response. Study approval was provided by the university's Institutional Review Board (IRB). Detailed demographics including race, gender, and computer competency can be found in the supplemental data.

Among BLVI participants, 9 reported being totally blind, 7 had light perception, 3 had low vision, and 1 could see shapes. Eleven participants had been blind since birth, while 9 acquired vision loss between ages 3 and 35.

*3.2.5Procedure*

Before starting the experiment, participants completed a demographic survey along with their consent form. Participants were then provided an email confirming they had Zoom installed for video conferencing, and an updated browser (Chrome, Edge, Safari, or Firefox) for running the study. When the study began, participants were provided a link to the study website. The study took around an hour and took place remotely over Zoom.

Participants were initially presented with the training map, which was one of the four maps used in the experimental conditions (see figure 2). They viewed the training map in multiple modalities (ITM and table for BLVIs; ITM, table, and visual for sighted participants). Participants were shown how to use each modality and answered 10 warm-up questions. The training session lasted 15 minutes, with participants allowed to stop early if desired. Participants were asked to answer at least one question in each modality.

After the training session, participants were shown the first experimental map. The map order and modality order were counterbalanced across participants. Each participant answered 15 open-ended questions for each map, with the option to skip questions (note: skipped questions were later scored as "incorrect"). Although skipped answers may bias tables unfairly, participants were allowed to (and did) skip questions in other modalities. Each question had a two-minute time-out for blind participants (although this timeout was dropped for sighted participants), and response time was recorded from when the question was asked to when participants said "final answer."

Participants answered 15 open-ended questions in each experimental condition divided into three types: five numeric, five geographic, and five numeric geographic. Map features were referred to as "states" (instead of features), and "tests with a T" or "deaths with a D" was verbally spoken to avoid confusion. Example questions for each type included: Numeric—"What state has the highest number of total tests?"; Geographic—"Which state is directly west of Santa Clara?"; Numeric Geographic—"Which state bordering the east coast has the highest number of deaths today?" All questions can be found in the supplemental material.

After answering the spatial questions for each condition, participants completed a NASA Task Load Index (TLX) measure of perceived workload, and provided qualitative feedback on their experience using the map. Finally, after all experimental conditions, participants shared their thoughts on the conditions, including general impressions, difficulties encountered, differences between conditions, and ratings of preference and pleasure for each condition on a scale of 1-5.

*3.2.6Analysis*

All statistical analyses were performed separately for blind and sighted participant groups. This is because they were subject to fundamentally different interaction constraints, prior experience, and experimental conditions. BLVIs relied on screen readers with varying experience levels, and were subject to fixed time limits, whereas sighted participants received more streamlined training and clarification, and were not time-constrained. These differences introduce systematic confounds related to modality, cognitive and temporal load, and tool familiarity that would violate assumptions of equivalence required for between-group statistical comparison, making within-group analyses more appropriate for interpreting results.

The dependent variables of performance, preference, and subjective evaluations were analyzed using statistical tests selected based on assumption testing conducted prior to each analysis. A significance alpha threshold of $p < .05$ was used throughout, with Bonferroni corrections applied where appropriate for multiple comparisons.

Prior to conducting statistical tests, assumptions of normality were evaluated using Shapiro-Wilk tests and the presence of outliers was assessed by identifying values beyond three standard deviations from the mean (results for each test can be found in the supplemental data). To ensure methodological rigor, both parametric and non-parametric tests were reported in the Performance analysis in the supplemental data as a robustness check, and the results were the same. Each result reports the test used.

To assess within-subject consistency across conditions, Pearson and Spearman correlations were calculated between conditions and question types for each participant group. To evaluate the internal validity of the experimental design, counterbalancing analyses were conducted to assess potential order effects and materials effects, revealing no significant order effects for blind (p = .828) or sighted (p = .232) participants, no significant materials effects for blind participants (p = .098), and although sighted participants showed significant materials effects (p = .004), full counterbalancing across conditions ensured these stimulus-specific differences did not compromise the validity of condition comparisons, thereby confirming the internal validity of the design.

Qualitative feedback was used to inform future iterations of the tools and help explain results.

*3.2.7Hypotheses*

Hypotheses are numbered to correspond with the sub-research questions above.

- H1.1a: The ITM will have significantly higher overall performance than the table for blind participants. This is because the table lacks the spatial information required for two-thirds of the questions.
- H1.1b: Both the ITM and visual map will have significantly higher overall performance than the table for sighted participants. This is because the table lacks the spatial information required for two-thirds of the questions.
- H1.2: The map conditions (ITM for blind participants; ITM and visual map for sighted participants) will show significantly higher performance than the table on Spatial Knowledge Geographic and Spatial Knowledge Numeric Geographic questions, but not on Spatial Knowledge Numeric questions (applying to both groups where the visual map was presented). This is because tables communicate numeric information comparably to maps but communicate no spatial information.
- H1.3: Sighted participants' accuracy on Spatial Knowledge Geographic and spatial knowledge numeric geographic questions will be significantly higher in the visual map condition than in the ITM condition. This is because sighted participants are familiar with visual maps, visual maps have had thousands of years of refinement, and sighted participants had only 15 minutes to learn the ITM condition, an interface first evaluated in 2019. The reason for testing this seperately is to evaluate if there is evidence supporting if the ITM serves the equivalent purpose to a visual map.
- H1.4a: Blind participants will report significantly higher confidence when using the ITM than the table. This is because the ITM provides spatial information that enables meaningful engagement with geographic questions, which should translate into greater self-reported confidence.
- H1.4b: Sighted participants will report significantly higher confidence for the visual map than for the ITM and the table, and higher confidence for the ITM than for the table. This is because confidence is expected to track the amount of spatial information communicated and participants' prior familiarity with each representation.
- H1.5a: Blind participants will report significantly higher overall cognitive workload for the ITM than for the table. This is because the ITM requires constructing and maintaining a spatial mental model from sequential auditory information, whereas the linear structure of tables is more familiar and imposes lower demand.
- H1.5b: Sighted participants will report cognitive workload ordered as ITM > visual > table, with significantly higher workload on the ITM than on the visual and table conditions. This is because the ITM conveys more information than the table and is less familiar than the visual map, increasing mental demand and effort.
- H1.6a: Blind participants will significantly prefer the ITM over the table. This is because the ITM communicates the spatial information that the table lacks, supporting more meaningful task engagement.
- H1.6b: Sighted participants will prefer conditions in the order visual map > ITM > table, with the visual map significantly preferred over both other conditions. This is because preference is expected to track the amount of spatial information communicated and sighted participants' lifelong familiarity with visual maps.

- H1.7: For both groups, within-subject performance on Spatial Knowledge Geographic questions will not be positively correlated between map conditions (visual, ITM) and the table condition, and may be negatively correlated. This is because tables lack spatial information, so participants with strong spatial reasoning on maps cannot apply that reasoning to tables, producing near-chance or floor-level table performance regardless of map ability.

## 4 RESULTS

In the following results, groups refer to blind and sighted participants; conditions refer to visual map, ITM (interactive text map), and table; and question types include Spatial Knowledge Numeric, Spatial Knowledge Geographic, and Spatial Knowledge Numeric Geographic.

### 4.1 Performance (H1.1a, H1.1b, and H1.2)

**Blind Participants.** A Wilcoxon signed-rank test was conducted to compare task performance accuracy between the ITM and table conditions for **Blind Participants.** There was a significant difference in accuracy scores, $V = 187.50$, $p = .002$, with participants demonstrating substantially higher accuracy when using the ITM ($M = 0.44$, $SD = 0.50$) compared to the table ($M = 0.23$, $SD = 0.42$). The null hypothesis was rejected, supporting H1.1a. The effect size was large ($r = 0.69$). These results indicate that the ITM representation significantly enhanced blind participants' ability to answer spatial questions compared to tables.

A Friedman test was conducted to examine the effect of question type on performance. The analysis revealed a statistically significant main effect of question type, chi-squared(2) = 27.11, $p < .001$. Post-hoc pairwise comparisons revealed significant differences between all three question types: Spatial Knowledge Geographic vs. Spatial Knowledge Numeric ($p = .001$), Spatial Knowledge Numeric Geographic vs. Spatial Knowledge Numeric ($p = .001$), and Spatial Knowledge Numeric Geographic vs. Spatial Knowledge Geographic ($p = .014$). This indicates that blind participants' performance varied substantially depending on the cognitive demands imposed by different question categories.

Stratified analyses examining the condition effect at each question type level revealed that the advantage of ITMs over tables was most pronounced for Spatial Knowledge Geographic and Spatial Knowledge Numeric Geographic questions, where spatial information is essential, whereas performance on Spatial Knowledge Numeric questions showed less differentiation between conditions. This pattern supports H1.2 for blind participants.

**Sighted Participants.** A Friedman test was conducted to examine the effect of presentation condition (visual map, ITM, table) on task performance accuracy for **Sighted Participants.** The analysis revealed a statistically significant main effect of condition, chi-squared(2) = 31.82, $p < .001$, Kendall's $W = 0.795$. Post-hoc pairwise Wilcoxon signed-rank tests with Bonferroni correction revealed the following patterns:

Participants achieved significantly higher accuracy on the ITM ($M = 0.72$, $SD = 0.45$) compared to the table ($M = 0.32$, $SD = 0.47$), $p < .001$, representing a mean difference of 0.40. Participants achieved significantly higher accuracy on the visual map ($M = 0.82$, $SD = 0.39$) compared to the table, $p < .001$, representing a mean difference of 0.50. A significant difference was also observed between the visual map and ITM conditions, $p = .019$, with visual presentation yielding higher accuracy by a margin of 0.10. The null hypothesis was rejected, supporting H1.1b.

A Friedman test revealed a significant main effect of question type on sighted participants' performance, chi-squared(2) = 34.33, $p < .001$. Questions requiring geographic information were more challenging than purely numeric questions (all $p < .001$), likely because geographic information was inaccessible in the table condition. Stratified analyses showed significant condition effects for Spatial Knowledge Geographic (chi-squared(2) = 28.61, $p < .001$) and Spatial Knowledge

Numeric Geographic questions (chi-squared(2) = 31.55, $p < .001$), but not for Spatial Knowledge Numeric questions ($p = .607$). Post-hoc comparisons indicated that visual maps and ITMs were equally effective for geographic questions ($p = .202$), and both substantially outperformed the table ($p = .0003$). This pattern supports H1.2 for sighted participants.

### 4.2 Visual vs. ITM Performance on Spatial Knowledge Geographic Questions (Sighted Participants) (H1.3)

A paired-samples t-test was conducted to compare sighted participants' performance on Spatial Knowledge Geographic questions between the ITM and visual map conditions. No statistically significant difference was observed, $t(19) = -1.91$, $p = .071$. Mean accuracy was comparable between the ITM ($M = 0.59$, $SD = 0.28$) and visual map ($M = 0.73$, $SD = 0.19$) conditions, with a mean difference of -0.14. The effect size (Cohen's $d = -0.43$) indicates a small-to-medium effect favoring the visual condition. We failed to reject the null hypothesis; H1.3 was not supported.

### 4.3 Subjective Confidence (H1.4)

**Blind Participants.** A paired-samples t-test was conducted to examine differences in self-reported confidence between the ITM and table conditions for **Blind Participants.** No statistically significant difference was observed, t(19) = -0.16, p = .878. Mean confidence ratings were comparable between the ITM (M = 3.25, SD = 1.12) and table (M = 3.30, SD = 1.26) conditions. This finding suggests that despite substantial differences in task performance across modalities, blind participants' subjective confidence remained relatively stable. We failed to reject the null hypothesis; H1.4a was not supported.

**Sighted Participants.** A Friedman test was conducted to examine the effect of presentation condition on self-reported confidence for **Sighted Participants.** The analysis revealed no statistically significant effect, chi-squared(2) = 5.63, p = .060. Participants reported comparable confidence levels across visual map (M = 4.15, SD = 0.59), ITM (M = 3.45, SD = 0.69), and table (M = 3.28, SD = 1.19) conditions, suggesting that subjective confidence was not strongly influenced by presentation modality. Post-hoc pairwise Wilcoxon tests with Bonferroni correction revealed a significant difference between the visual and ITM conditions (p = .004), but no significant differences for other comparisons. We failed to reject the null hypothesis at the overall level; H1.4b was only partially supported (visual vs. ITM pairwise difference).

### 4.4 NASA Task Load Index (Cognitive Workload) (H1.5)

**Blind Participants.** Cognitive workload was assessed using the NASA Task Load Index (NASA-TLX). A paired-samples t-test comparing overall workload between presentation conditions revealed a statistically significant difference, t(19) = 2.48, p = .023. Blind participants reported lower cognitive workload when using the table (M = 27.40, SD = 8.74) compared to the ITM (M = 34.10, SD = 7.69), representing a medium effect (Cohen's d = 0.55). The null hypothesis was rejected, supporting H1.5a.

This finding indicates that while the ITM provides richer spatial information and supports better task performance, it imposes greater cognitive demands on blind users. The higher workload associated with the ITM likely reflects the additional cognitive resources required to construct and maintain spatial mental models from sequential auditory information, compared to the more familiar linear structure of tabular data. Subscale analysis revealed that Mental Demand, Temporal Demand, and Effort were all significantly higher for the ITM condition, with Effort showing the largest effect (Cohen's $d = 0.76$; details in supplemental material).

**Sighted Participants.** Cognitive workload for sighted participants was examined across all three presentation conditions using a repeated measures ANOVA. The overall ANOVA was non-significant, suggesting that overall cognitive workload was statistically comparable across presentation modalities for sighted users (table: M = 16.00; visual: M =

17.00; ITM: M = 27.30). Subscale analysis revealed significant differences in Mental Demand, Physical Demand, and Effort, all highest for the ITM condition (details in supplemental material). Frustration was significantly higher for ITM compared to visual only. We failed to reject the null hypothesis at the overall level; H1.5b was only partially supported at the subscale level.

### 4.5 Preference (H1.6)

**Blind Participants.** A Wilcoxon signed-rank test was conducted to examine differences in subjective preference ratings between the ITM and table conditions for **Blind Participants.** The analysis revealed a statistically significant difference, V = 143, p = .002. Blind participants expressed a significantly stronger preference for the ITM (M = 4.10, SD = 1.07) compared to the table (M = 1.95, SD = 1.43), with a large effect size (r = 0.71). This finding indicates that blind users not only performed better with the ITM but also subjectively preferred this modality over tables. The null hypothesis was rejected, supporting H1.6a.

**Sighted Participants.** Subjective preference ratings were analyzed using a Friedman test to examine whether sighted participants demonstrated differential preferences across conditions. The analysis revealed a statistically significant main effect of condition, chi-squared(2) = 28.46, p < .001. Mean preference ratings were 4.60 for the visual map, 2.55 for the ITM, and 1.65 for the table. Post-hoc pairwise Wilcoxon tests with Bonferroni correction revealed that the visual map was significantly preferred over both the ITM (p = .001) and the table (p < .001). The difference between ITM and table preferences approached but did not reach significance (p = .164). These findings indicate that sighted participants' subjective preferences varied systematically across presentation modalities, with a clear preference hierarchy of visual map > ITM > table. The null hypothesis was rejected, supporting H1.6b (with the caveat that the ITM vs. table pairwise comparison did not reach significance).

### 4.6 Within-Subject Correlations (H1.7)

Within-subject correlations were examined to assess consistency of individual performance across conditions. For blind participants, overall correlations between ITM and table conditions were not significant (r = -0.05, p = .846), though significant negative correlations emerged for geographic questions (r = -0.51, p = .021; rho = -0.45, p = .045), suggesting that blind participants who performed well on geographic questions using the ITM performed worse on the table. For sighted participants, correlations between conditions were low and non-significant (all p > .13), with several correlations uncomputable due to zero variance in negative table performance on geographic questions. These patterns reflect participants were unable to obtain spatial information from the table. The null hypothesis was rejected for blind participants on geographic questions, supporting H1.7.

### 4.7 Summary of Key Findings

In summary, the results demonstrate that: (1) Both visual and ITMs significantly outperformed tables in overall task accuracy for both blind and sighted participants, with large effect sizes; (2) The performance advantage of maps over tables was particularly pronounced for Spatial Knowledge Geographic and Spatial Knowledge Numeric Geographic questions, whereas Spatial Knowledge Numeric questions showed minimal condition effects; (3) No significant difference was observed between visual and ITM conditions for sighted participants on Spatial Knowledge Geographic questions, suggesting comparable effectiveness in conveying spatial information; (4) Blind participants reported significantly higher cognitive workload for the ITM compared to the table, though they achieved substantially better performance with the ITM; (5) Both blind and sighted participants preferred the visual and ITMs over tables, with blind participants showing a strong preference for the ITM and sighted participants preferring the visual map; and (6) Subjective confidence did not significantly differ across conditions, despite objective performance differences.

### 4.8 Qualitative Analysis

Both blind and sighted participants were very clear in their comments that the ITMs and visual maps provided more information than the table. The code frequency for "Table Map Geographical Information Accessibility" was 189 across both blind and sighted participants, which was the second most frequent code, just behind ITM Design (200). This code specifically focused on the lack of geographic accessibility in the Table condition. Comments comparing the ITM and Table conditions had 90 codes. One participant commented: "ITM has, I guess, the ability to figure out the geographical data, whereas the table map doesn't." (Sighted participant P1). The vast majority of geographic questions were answered with a"skip" on the Table condition. Participants did appreciate the ease of use when on the Table condition. The conclusion is that a table is better than the visual and ITM at answering numeric questions, but is unable to help with geographic information.

Another set of notable codes focused on ITM Design (frequency 200), and ITM Difficulty with Navigation (frequency 51). For BLVIs, many of the issues were with using their screen reader. Since Audiom is a web component that requires users to find a particular button on a page, and since the study required the BLVIs to have three tabs open, there was often confusion about how to get to Audiom in the first place. That is an issue of the experiment design. BLVIs also had difficulty navigating the table with standard screen reader commands. Additionally, participants sometimes pressed incorrect keys, unbeknownst to the researcher, and it took the researcher a while to figure out what was happening, and why the commands were not responding as desired.

Another aspect of these codes had to do with the experience of participants once they actually started using the ITM experience of Audiom. BLVIs found the extra geographic information more difficult to sort through than the table. Participants mentioned that it was more difficult for them to visualize the shapes than it would be if they were touching a tactile graphic, especially when only using the text element. Participants pointed out that adding audio to the text made the ITM geographic experience much better, as they could quickly move around the map without waiting for the name and other values to read through speech. The ITM commands were also difficult for participants to remember. Although the participants learned seven key commands (four of which were arrow keys), some still found the learning curve steep and commented they did not remember how to use Audiom while performing the study condition.

Features participants requested to improve the ITM interface included: ability to use first-letter navigation on the tabular view of Audiom; make the key commands more logical; make it easier to follow the border of map features; mention if the state is landlocked; add new sounds to the map; switch to using other measurement units other than meters; ability to turn off the sounds to make a text only experience; jump from one border to another border of a feature; a command listing bordering features; quiet the sound when moving to another location to prevent overlap; ability to adjust the volume of sounds within Audiom; having both the tabular view and geographic view next to each other visually; have a tutorial one could refer to at any time; and sighted participants wanted the values read aloud in the tabular view of Audiom.

One technique sighted participants often used to help comprehend the ITM was visually redrawing the ITM with a pencil on a piece of paper as they explored it. This allowed them to understand the geographic layout once in the ITM, then use their vision to answer future spatial questions.

Participants mentioned that one of the most confusing aspects of the experience was the ambiguity around what was considered a border state versus a landlocked state. Because the shapes were rotated and shifted, this created odd gaps between the different states similar to geological features such as bays, sounds, peninsulas, and rivers. Although this is realistic, it created ambiguity in answers (e.g., is a feature bordering a "bay" considered on the coast?). To correct for these ambiguities, several answers were accepted for those questions. Despite this, future work should focus on removing ambiguities by removing gaps between features, despite this being less "realistic". Because there were 20 features, accepting two similar answers (e.g., "Johnson City, Bigford, and Sunny Bluffs" and "Freemont, Johnson City, Bigford, and Sunny Bluffs") was not deemed to have significantly impacted the nature of the findings.

## 5 DISCUSSION

### 5.1 Tables Fail to Communicate Spatial Information

The primary conclusion of this study is that tables, as currently used as an accessible map lacking a geographic column, are significantly worse than either visual maps or ITMs at communicating geographic information ($p < .001$) these tables would be unable to communicate the MEP variables of map features [Biggs et al., 2026]. Multiple lines of evidence converge on this conclusion.

First, responses in the table condition showed a uniform distribution ($M = 0.32$, $SD = 0.47$), likely because two-thirds of questions required spatial information and correct answers were presumably due to chance. Second, most participants requested to skip spatial questions when using the table. Third, the negative correlation ($r = -0.51$) between ITM and table performance on geographic questions for blind participants demonstrates that tables fundamentally lack spatial information. Participants with strong spatial reasoning (evidenced by high ITM performance) showed correspondingly poor table performance—reflecting the impossibility of answering spatial questions without spatial information. Fourth, BLVI participants had equal confidence across ITM and table conditions (M = 3.25 vs. 3.30, p = .878), despite the ITM producing nearly double the accuracy (M = 0.44 vs. 0.23). This suggests BLVIs may not understand they are missing geographic information, even when it is critical for a correct response.

These findings are consistent with prior work, though direct comparisons require accounting for methodological differences. For instance, Zhao et al. reported higher accuracy (95% vs. 20% for tables) than the present study (44% for BLVIs on ITMs), but their interface omitted shape, size, orientation, distance, and topological relationships, included 109 minutes of training versus 15 minutes here, and imposed no time limits [Zhao et al., 2008]. Smelcer et al. further demonstrated that even tables enhanced with spatial descriptions of adjacent features remained significantly less effective than visual maps for distance-based questions [Smelcer and Carmel, 1997]. Despite these methodological variations, all studies converge on the same conclusion: tables, especially when lacking geographic information, are dramatically inferior to map interfaces for spatial questions. This cumulative evidence, combined with results from the MEP Framework, provides sufficient justification to remove tables from recommended text map alternatives in digital accessibility guidelines [Colorado OIT-GIS, 2024; McCall, 2024; Sloan, 2020].

### 5.2 Preference and Cognitive Load

Preference for each condition correlated with the amount of spatial information it communicated: visual map > ITM > table for sighted participants, and ITM > table for BLVIs. Sighted participants likely preferred the visual map because visual maps have had thousands of years of refinement [Arundel and Li, 2021; Longley et al., 2011c], whereas Audiom was first evaluated in 2019 [Biggs et al., 2019]. BLVIs significantly preferred the ITM over the table despite the table being more familiar, because the ITM communicated spatial information absent from the table.

NASA-TLX scores followed the pattern: ITM > visual > table. The ITM scores ($M = 34.1$ for BLVIs; $M = 27.3$ for sighted participants) fall on the lower end of the 0–100 scale, indicating room for improvement but not problematic workload. Mental demand and effort were the highest subscales for both the ITM and visual conditions, likely reflecting the increased information these interfaces provided. The apparent paradox of higher workload alongside better performance reflects information access: the table's lower workload simply meant participants could not engage with the two-thirds of questions requiring spatial information. Lower workload is not inherently desirable if it comes at the cost of information access.

### 5.3 Theoretical Implications

The finding that there was no significant difference between the ITM and visual conditions for sighted participants on geographic question performance suggests that the ITM fully conveys the spatial elements of a map and can function as a non-visual geographic map representation. This is not because the ITM showed equivalency, but because participants were able to accurately answer spatial questions with the ITM at a similar level to the visual map. Future work should investigate true equivalency with a larger sample size, but these results are promising. The difficulty BLVIs demonstrated when answering numeric geographic questions may reflect the cognitive load involved in switching between interfaces (e.g., geographic view versus tabular view in Audiom). Additionally, both mental demand and effort were higher for the ITM and visual conditions than for the table among sighted participants, yet mental demand and effort did not differ significantly between the ITM and visual conditions. The overall NASA-TLX score was significantly different between the visual and ITM conditions for sighted participants ($p < .05$). This suggests that the ITM and visual representations engage similar cognitive processes, and it is expected that both map types would have higher workload than a table due to the greater amount of information they convey.

### 5.4 Implications for Design

Participant feedback revealed several opportunities to improve the ITM interface. Requested features included first-letter navigation in the tabular view, more intuitive key commands, border-following functionality, automatic landlocked state detection, adjustable sound volume, and the ability to view geographic and tabular views simultaneously.

An unexpected finding concerned table navigation preferences. Although the Audiom table interface was designed to require only arrow keys (similar to Excel), blind participants often preferred using their familiar screen reader table navigation commands (Ctrl+Alt+Arrow keys). This preference for familiar interaction patterns, even when objectively less efficient, highlights the importance of supporting multiple navigation paradigms. However, this design choice has scalability implications: standard screen reader table navigation becomes impractical beyond approximately 250 rows, after which pagination or dynamic loading becomes necessary. Since thematic maps may contain thousands of features, future implementations should balance familiarity with scalability.

The technique observed among sighted participants (redrawing the ITM on paper while exploring) suggests that multiple sensory modalities makes the interface more usable to everyone, including able-bodied users who could use the ITM interface effectively.

It is important to note that although this paper sometimes uses "audio" shorthand for the Audiom condition, and participants in this study interacted with the interface only through audio output, the interface is fundamentally an ITM. Outside this study, sighted users can read all labels through on-screen text, and the text is sent to the user's screen reader, making it accessible on a braille display. The audio and visual interfaces are fully integrated on the Audiom platform [Welcome to XR navigation!]. This integration is critical for WCAG compliance. Rather than focusing solely on improving the sonification experience, all content is constrained to text-based presentation. The audio and visual experiences are supplemental to the text experience.

### 5.5 Policy Implications

Some digital accessibility practitioners argue that only the “primary purpose” of a map needs to be conveyed through a text description [KMSOC, 2024; Sloan, 2020; UniDescription academy]. However, WCAG requires that text descriptions serve an “equivalent purpose” to non-text content [World Wide Web Consortium]. All maps communicate “generalized spatial information and relationships” [Biggs et al., 2026; Lapaine et al., 2021]; if that spatial information is unnecessary, a different representational method should be used for everyone. If sighted users are given a map because geographic representation is required for communication (see Figure 3), then WCAG requires that BLVIs be provided with a geographic map that fully communicates geographic information as well [Biggs et al., 2026; World Wide Web Consortium].

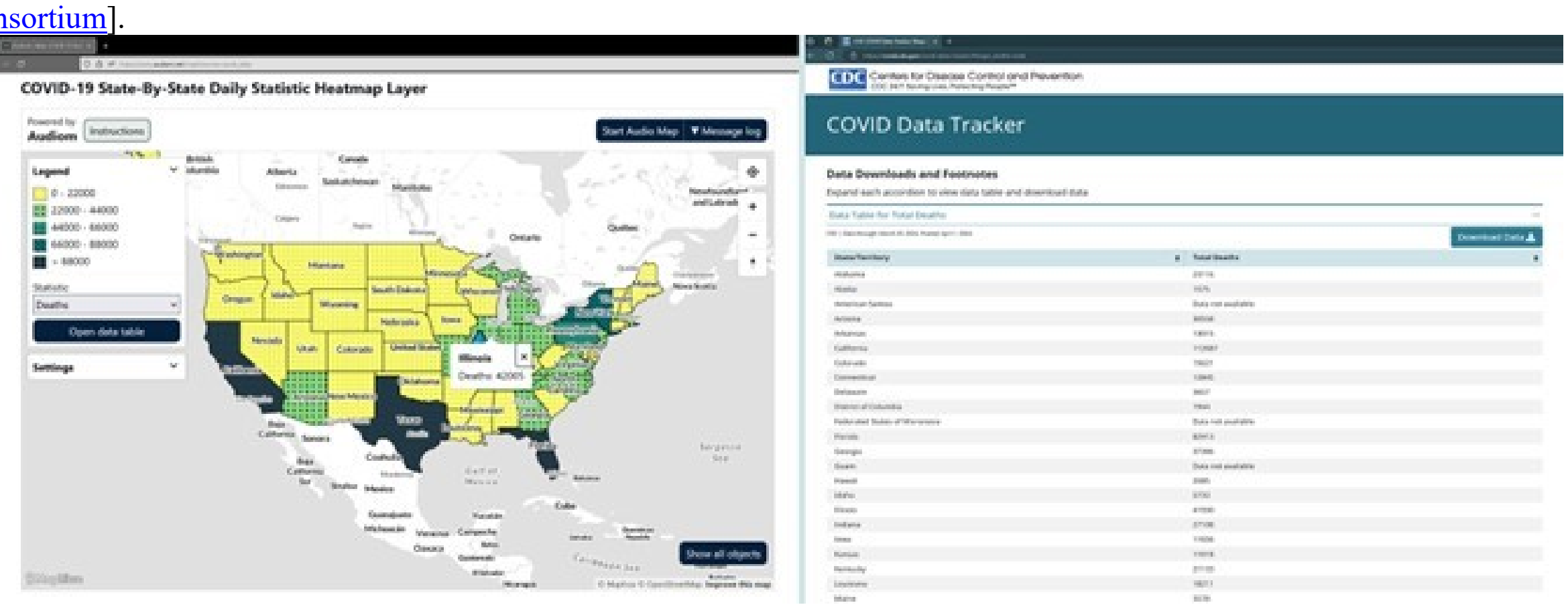

Figure 3: A comparison of an Audiom choropleth map showing COVID statistics compared to the CDC website showing a table of data

Note: Here is a link to an interactive example of the table from the CDC website and the Audiom map (Opens a New Tab).

CDC source [United states COVID-19 cases, deaths, and laboratory testing (NAATs) by state, territory, and jurisdiction].

International accessibility legislation, including the European Union’s directive [European Commission, 2017], the United Kingdom’s regulations [The public sector bodies (websites and mobile applications) (no. 2) accessibility regulations 2018, 2018], and the Accessible Canada Act [Government of Canada, 2025], have historically made explicit exceptions for digital thematic maps, likely because accessible alternatives were not available. This is no longer the case. Section 508 in the United States requires that “If there are technically acceptable solutions available in the marketplace, you must select one of those solutions” [Buy accessible products and services], and the U.S. Department of Justice’s ADA Title II rule taking effect in April 2027 explicitly mentions that “maps” must be WCAG AA compliant [U.S. Department of Justice Civil Rights Division, 2024]. As demonstrated by the MEP Framework [Biggs et al., 2026], the systematic WCAG map evaluation [Biggs et al., 2025], and the empirical results presented here, it is now possible to make digital thematic maps fully usable to BLVIs. These legislative exceptions should be reconsidered, as they prevent BLVIs from accessing public health, weather, demographic, and other geo-referenced data. A potential objection is that ITMs are more difficult to create than tables; however, ITMs like Audiom use the same GeoJSON data that powers visual maps, requiring no additional data creation [Working Group Geographic JSON (geojson), 2016]. Legacy maps can also be made accessible provided they can be vectorized into standard GeoJSON or other vector format.

### 5.6 Limitations

Several methodological limitations should be noted. The table condition was designed to reflect current accessibility practice rather than to represent the strongest possible table alternative. The 2-minute timeout imposed on blind participants but not sighted participants likely contributed to lower overall accuracy in the BLVI group and may have elevated the temporal demand subscale of the NASA-TLX; this protocol difference, along with slight procedural modifications between groups, limits direct between-group comparisons and should be standardized in future studies. Screen reader interaction difficulties reported by some BLVI participants may have reduced their absolute performance levels, though because all statistical analyses were conducted within-group, these usability issues do not affect the primary conclusions regarding condition differences. Additionally, this study relied exclusively on verbal question-answering to assess spatial knowledge acquisition; non-verbal methods such as map reproduction tasks (e.g., using tactile materials, sketching, or digital reconstruction) may capture aspects of spatial mental models that verbal responses cannot fully express and should be incorporated in future research.

The sample of 20 blind and 20 sighted participants limits the power of between-group comparisons. The blind participant group ($M$ = 41.2 years, range = 21-68) was significantly older than the sighted group ($M$ = 20.3 years, range = 18-24), $t(21.5) = 7.39$, $p < .001$; however, age was not significantly correlated with performance within either group (blind: $r = -0.02$, $p = .929$; sighted: $r = 0.00$, $p = .993$), suggesting age differences do not confound the within-group analyses. Blind participants reported high computer usage skill ($M$ = 4.30, $SD$ = 0.66 on a 5-point scale), and self-reported computer skill was not significantly correlated with overall performance (Spearman *rho* = -0.15, $p = .523$), indicating that success with the ITM was not dependent on prior technology expertise, though future research should include participants with a wider range of technology experience [Satterfield et al., 2025]. No separate measure of spatial ability was included; prior work found no significant correlation between self-reported spatial ability and tactile map task performance [Brock et al., 2015], but future research could examine whether spatial ability moderates ITM effectiveness.

All participants were first-time Audiom users with only 15 minutes of training, so ITM performance may represent a lower bound that extended practice could improve. A significant materials effect was observed for sighted participants ($p = .004$), indicating the three experimental maps varied in difficulty; however, full counterbalancing across conditions ensures this does not compromise condition comparisons. The spatial knowledge questions were not designed to explicitly capture all variables of the MEP Framework because the framework was published after data collection had begun, and the maps contained only fictional irregular polygons. Additionally, the maps should be improved by removing the spaces between features and smoothing out borders; although the current features are more realistic, the ambiguities they introduce (e.g., whether a feature bordering a gap is coastal) made some questions harder to answer definitively. Future studies should standardize protocols across groups, incorporate non-verbal spatial assessment, use a larger and more diverse sample, and explicitly evaluate all MEP Framework variables.

### 5.7 Generalizability and Future Work

Fictional maps with unfamiliar place names were used to control for prior geographic knowledge. This study focused exclusively on choropleth maps with non-overlapping polygons; other thematic map types (e.g., proportional symbol maps, dot density maps, isopleth maps) may present different challenges, particularly with overlapping geometries. The MEP Framework suggests any thematic map should be representable through interactive text, but empirical validation across map types remains for future work [Types of thematic maps, 2017]. Future work should also investigate using points and lines in addition to polygons. The 20-feature maps represent moderate complexity; scalability to larger feature collections requires further investigation.

Future work should address several areas: (1) true equivalency testing between ITM and visual maps with a larger sample size, using properly powered equivalency testing adapted from clinical trial methodology to determine whether ITM and visual map performance fall within a predefined equivalence margin [Lakens et al., 2018], with questions that explicitly target all MEP Framework variables, and potentially evaluating whether ITM users can pass standardized geography assessments (e.g., the NAEP geography assessment [The Institute of Education Sciences, 2026]) at rates comparable to visual map users; (2) implementation and evaluation of participant-recommended features (first-letter navigation, border-following, automatic landlocked detection, simultaneous geographic and tabular views); (3) longitudinal studies examining how performance and workload change with extended practice; (4) explicit tests with points and lines to validate generalization beyond polygons; and (5) non-verbal spatial assessment methods (e.g., map reproduction using tactile materials) to complement verbal measures.

## 6 CONCLUSION

This evaluation yields two principal findings. First, tables traditionally used for accessibility, fail to communicate spatial information and therefore do not serve an equivalent purpose to a visual or interactive text map. Across both sighted and blind participants, map representations significantly outperformed tables on geographic questions, while participants overwhelmingly preferred maps and qualitative feedback confirmed that traditional tables simply could not support spatial reasoning. Tables should no longer be considered an acceptable alternative to a geographic map. Second, results suggest that a WCAG-compliant ITM, such as Audiom, can communicate spatial information. Sighted participants showed no significant performance difference between visual and ITM conditions on geographic questions, which suggests a larger powered study may find equivalence between conditions. These findings provide empirical grounds for reconsidering international legislative exceptions that exempt digital thematic maps from accessibility requirements [European Commission, 2017; Government of Canada, 2025; The public sector bodies (websites and mobile applications) (no. 2) accessibility regulations 2018, 2018]. Future non-visual map research should adopt WCAG compliance as a prerequisite for adoption arguments and use the MEP Framework as a shared baseline for comparing map representations. Future work on the Audiom interface should explore additional comparison methods (e.g., map reproduction, navigation tasks), extend evaluations to audio descriptions and multi-user ITMs, and improve the table condition. Future text map evaluations should perform a similar comparison between audio descriptions or Multi User Domain ITMs and visual maps, as the MEP Framework suggests they should also serve an equivalent purpose to a visual map. Establishing formal equivalence between ITM and visual map conditions through properly powered testing remains a critical next step.

## 7 SUPPLEMENTAL MATERIAL

### 7.1 Platform with conditions

The site with exact conditions used in this study can be found at:

https://xrnavigation.github.io/tvm-study/

### 7.2 Code

The code for the study site can be found at:

https://github.com/xrnavigation/tvm-study

### 7.3 R Code an Anonymized Raw Data

The complete R code, compiled R-Markdown HTML report, and raw participant data can be found at:

https://doi.org/10.3886/E247550V1

## 8 ACKNOWLEDGMENTS

We would like to thank all the amazing research assistants who worked on this project.

### 8.1 Funding

This project was funded under NIH Grants No. 1 R44 EY036316-01A1 and 1R41EY034411-01A1, as well as NIDILRR Grant No. 90REGE0018.

### 8.2 Conflicts of Interest

Brandon Biggs is the CEO of XR Navigation, the company commercializing Audiom, the platform used in this study.